\documentclass[a4paper,11pt]{article}
\pdfoutput=1 % if your are submitting a pdflatex (i.e. if you have
\usepackage{jcappub} % for details on the use of the package, please
\usepackage{xcolor}
\usepackage[T1]{fontenc} % if needed
\newcommand{\apj}{Astrophys. J.}
\newcommand{\mnras}{Mon. Not. R. Astron. Soc.}
\newcommand{\aap}{Astron. Astrophys.}

\newcommand{\araa}{Ann. Rev. Astron. Astrophys.}
\newcommand{\apjl}{Astrophys. J.}
\newcommand{\pasj}{Publ. Astron. Soc. Jpn.}
\newcommand{\jcap}{J. Cosmol. Astropart. Phys.} 
 
\newcommand{\aaps}{Astron. Astrophys. Sup.} 
\newcommand{\apss}{Astrophysics and Space Science} 
\newcommand{\nat}{Nature} 
\newcommand{\prd}{Phys. Rev. D}
\newcommand{\prc}{Phys. Rev. C}

\title{Sagittarius A* as an Origin of the Galactic PeV Cosmic Rays?}

\author[a]{Yutaka Fujita}
\author[b]{Kohta Murase}
\author[b]{Shigeo S. Kimura}

\affiliation[a]{Department of Earth and Space Science, Graduate School
 of Science, Osaka University, Toyonaka, Osaka 560-0043, Japan}
\affiliation[b]{Center for Particle and Gravitational Astrophysics;
 Department of Physics; Department of Astronomy \& Astrophysics, The
 Pennsylvania State University, University Park, PA 16802, USA}

\emailAdd{fujita@vega.ess.sci.osaka-u.ac.jp}
\emailAdd{murase@psu.edu}
\emailAdd{szk323@psu.edu}

\abstract{Supernova remnants (SNRs) have commonly been considered as a
source of the observed PeV cosmic rays (CRs) or a Galactic PeV particle
accelerator ("Pevatron"). In this work, we study Sagittarius A* (Sgr
A*), which is the low-luminosity active galactic nucleus of the Milky
Way Galaxy, as another possible canditate of the Pevatron, because it
sometimes became very active in the past. We assume that a large number
of PeV CRs were injected by Sgr A* at the outburst about $10^7$ yr ago
when the Fermi bubbles were created. We constrain the diffusion
coefficient for the CRs in the Galactic halo on the condition that the
CRs have arrived on the Earth by now, while a fairly large fraction of
them have escaped from the halo. Based on a diffusion-halo model, we
solve a diffusion equation for the CRs and compare the results with the
CR spectrum on the Earth. The observed small anisotropy of the arrival
directions of CRs may be explained if the diffusion coefficient in the
Galactic disk is smaller than that in the halo. Our model predicts that
a boron-to-carbon ratio should be energy-independent around the knee,
where the CRs from Sgr~A* become dominant. It is unlikely that the
spectrum of the CRs accelerated at the outburst is represented by a
power-law similar to the one for those responsible for the gamma-ray
emission from the central molecular zone (CMZ) around the Galactic
center.}
\begin{document}
\maketitle
\flushbottom

\section{Introduction}
\label{sec:intro}

The origin of cosmic rays (CRs) has been discussed for a long time.
Supernova remnants (SNRs) are the most popular candidate for CR protons
below the knee (at $\sim 10^{15.5}$~eV $\approx 3$~PeV) and CR nuclei
below the second knee (at $\sim100$~PeV).  In fact, X-ray observations
have revealed that electrons with energies of $E\sim 100$~TeV are
actually accelerated at SNRs~\cite{1995Natur.378..255K} and gamma-ray
observations have indicated that protons are also
accelerated~\cite{2010ApJ...710L.151T,2013Sci...339..807A}. Moreover,
efficient CR acceleration at SNRs is now supported by state-of-art
numerical simulations~(e.g.,
\cite{2013APh....43...56B,2014ApJ...794...46C,2014ApJ...794...47C,2015ApJ...802..115K}).
However, while the SNRs have been believed to accelerate protons up to
$\sim$~a~few~PeV~\cite{2002cra..book.....S,2015arXiv151007042C}, none of
them have been identified as PeV accelerators of CR protons
(``Pevatrons'').  Their gamma-ray spectra do not extend without a cutoff
only up to a few tens of
TeV~\cite{2011ApJ...730L..20A,2013APh....43...71A}.

In Ref.~\cite{2015PhRvD..92b3001F} (Paper~I; see also
Ref.~\cite{2005Ap&SS.300..255A,2007ApJ...657L..13B,2011ApJ...726...60C,2016arXiv160408301G,2016arXiv160408791C,2016arXiv160901069L}),
we argued that CR protons with $\gtrsim$~TeV are accelerated in the
active galactic nucleus (AGN) of the Milky Way Galaxy or Sagittarius A*
(Sgr~A*). This is because Sgr~A* is a low luminosity AGN (LLAGN), where
CR acceleration may occur in its radiatively inefficient accretion flow
(RIAF) and/or a vacuum gap in the black hole
magnetosphere~\cite{2007ApJ...659.1063K,2015ApJ...806..159K,2015PhRvL.114f1101H,2016PhRvL.116g1101M,2016ApJ...822...88K,2016A&A...593A...8P}.
While the level of activity of Sgr~A* is very low at the present,
observations of X-ray echos indicated that Sgr~A* had been much more
active more than 50~yrs
ago~\cite{1996PASJ...48..249K,2000ApJ...534..283M,2006PASJ...58..965T,2013PASJ...65...33R}.
We showed that the CR protons accelerated at Sgr~A* during that active
period escape into the interstellar space and part of them plunge into
the central molecular zone (CMZ), which is a dense gas ring surrounding
Sgr~A*~\cite{2015PhRvD..92b3001F}. These CRs interact with the molecular
gas in the CMZ and generate diffuse multi-TeV gamma-rays. The predicted
gamma-ray flux is shown to be consistent with that obtained with the
High Energy Stereoscopic System (HESS) \cite{2006Natur.439..695A}.  If
this scenarios is correct, it means that Sgr~A* had kept this level of
activities for at least $\sim 10^4$--$10^5$~yrs, which is the
diffusion time of those CRs in the CMZ~\cite{2015PhRvD..92b3001F}. CR
electrons might also have been accelerated through those
activities~\cite{2015JCAP...12..005C}. We note that CR protons may also
be produced in extragalactic LLAGNs, which can explain the extragalactic
neutrino flux observed with
IceCube~\cite{2013PhRvL.111b1103A,2013Sci...342E...1I,2014PhRvL.113j1101A,2015PhRvD..91b2001A}.
Interestingly, this model can account even for the 10-100~TeV
data~\cite{2015ApJ...806..159K,2016PhRvL.116g1101M}.

The activities of Sgr~A* more than $\sim 10^4$--$10^5$~yrs ago are not
well constrained~\cite{2002A&A...389..252C,2014IAUS..303..333P}. The
existence of the Fermi bubbles (FBs) suggests that there were even
stronger activities at the Galactic center
\cite{2010ApJ...724.1044S,2011PhRvL.106j1102C,2012MNRAS.423.3512C,2015ApJ...814...93S}. If
they were created by a short-term violent activity of Sgr~A*, it
happened $\sim 1$--$10$~Myr ago
\cite{2011MNRAS.415L..21Z,2012ApJ...756..181G,2012ApJ...761..185Y,2012MNRAS.424..666Z,2013ApJ...775L..20F,2014ApJ...789...67F}.
If a huge amount of CRs were injected by Sgr~A* at that outburst, they
may have filled in the Galactic halo and some of them may have
re-entered the Galactic disk \cite{2016arXiv160708862T}.  Since we do
not know much about the diffusion coefficient for PeV CRs in the
Galactic halo, a fraction of the CRs injected by Sgr~A* have not escaped
from the Galaxy if the coefficient has an appropriate value. In this
work, we study the possibility that the energies of those CRs are around
the knee and some of them have arrived on the Earth
\cite{1983JPhG....9.1139G}. Although such a single source scenario
sounds extreme, surprisingly, it has not been ruled out. In fact, those
CRs cannot be ignored because the total CR energy injected by the
outburst is $\sim 3\times 10^{54}\rm\: erg$ (see
eq.~\ref{eq:Lptot}). This is comparable to the total energy of CRs
accelerated by SNRs in the Galaxy, $E_{\rm CR,SNR}=\epsilon_{\rm p}
E_{\rm SN} R_{\rm SN} t_{\rm diff}\sim 10^{55}\rm\: erg$, where
$\epsilon_{\rm p}\sim 0.1$ is the CR acceleration efficiency, $E_{\rm
SN}\sim 10^{51}\rm erg$ is the energy released by one supernova
explosion, $R_{\rm SN}\sim 0.01\rm\: yr^{-1}$ is the occurrence rate of
supernova explosion in the Galaxy, and $t_{\rm diff,G}\sim 10^7$~yr is
the typical diffusion time of CRs in the Galaxy.

This paper is organized as follows.  In section~\ref{sec:model}, we
describe our model for the acceleration of CRs at Sgr~A* and their
diffusion in the Galactic halo. In section~\ref{sec:result}, we show the
results of our calculations for the fiducial model. In
section~\ref{sec:disc}, we discuss related topics such as the
boron-to-carbon (B/C) ratio and high-energy CRs above the knee. In
section~\ref{sec:CMZ}, we discuss the gamma-rays from the Galactic
center. In section~\ref{sec:power}, we study the case where CR spectrum
in the Galactic halo is similar to that around the Galactic
center. Finally, section~\ref{sec:conc} is devoted to conclusions. We
consider protons as CRs unless otherwise mentioned.

\section{The Model}
\label{sec:model}

\subsection{Outburst component}

\subsubsection{CR acceleration at Sgr~A*}
\label{sec:CRacc}

The mechanism of CR acceleration at Sgr~A* is not well-known especially
when Sgr~A* bursts. The CRs may be accelerated in jets launched from
Sgr~A* when Sgr~A* becomes active. In this study, however, we assume
that the CRs are stochastically accelerated in a RIAF in Sgr~A*,
although we do not rule out other acceleration mechanisms (see
sections~\ref{sec:high} and~\ref{sec:power}).

Our acceleration model is a simple one-zone model,
which is basically the same as that in
Refs.~\cite{2015ApJ...806..159K,2015PhRvD..92b3001F}. The spectrum of
the accelerated CRs can be derived as follows. First, the typical energy
of the CR protons, $E_{\rm p,eq}$, is evaluated by equating their
acceleration time to their escape time from the RIAF. The result is
\begin{eqnarray}
\label{eq:geq}
 \frac{E_{\rm p,eq}}{m_{\rm p} c^2} &\sim& 1.4\times 10^5 
\left(\frac{\dot{m}}{0.01}\right)^{1/2}
\left(\frac{M_{\rm BH}}{1\times 10^7\: M_\odot}\right)^{1/2}\nonumber\\
& &\times \left(\frac{\alpha}{0.1}\right)^{1/2}
\left(\frac{\zeta}{0.1}\right)^3
\left(\frac{\beta}{3}\right)^{-2}
\left(\frac{R_{\rm acc}}{10\: R_S}\right)^{-7/4}\:,
\end{eqnarray}
where $m_{\rm p}$ is the proton mass, $\dot{m}=\dot{M}/\dot{M}_{\rm
Edd}$ is the gas accretion rate ($\dot{M}$) toward the supermassive
black hole (SMBH) normalized by the Eddington accretion rate
($\dot{M}_{\rm Edd}$), $M_{\rm BH}$ is the mass of the SMBH, $\alpha$ is
the alpha parameter of the accretion flow, $\zeta$ is the ratio of the
strength of turbulent magnetic fields to that of the non-turbulent
magnetic fields, $\beta$ is the plasma beta parameter, $R_{\rm acc}$ is
the typical radius where particles are accelerated, and $R_S$ is the
Schwarzschild radius of the SMBH.  The Eddington accretion rate is given
by $\dot{M}_{\rm Edd} = L_{\rm Edd}/c^2$, where $L_{\rm Edd}=1.26\times
10^{38}\: (M_{\rm BH}/M_\odot)\:\rm erg\: s^{-1}$ is the Eddington
luminosity. As fiducial parameters, we adopt $\alpha=0.1$, $\zeta=0.15$,
$\beta=3$, and $R_{\rm acc}=10\: R_S$, which are close to the values
that reproduce the IceCube neutrino observations
\cite{2015ApJ...806..159K}. We here choose $\zeta$ that is three times
larger than that in Paper~I, which will allow us to match the CR
spectrum on the Earth with the observations (see later). The accretion
rate is determined so that it is large enough to produce a required
amount of CRs and create the FBs but is small enough to form a
RIAF. Previous studies have shown that an accretion flow becomes a RIAF
when $\dot{m}\lesssim 0.1$ \cite{2014ARA&A..52..529Y}\footnote{The
Eddington luminosity is defined as $\dot{M}_{\rm Edd}=10\: L_{\rm
Edd}/c^2$ in Ref.~\cite{2014ARA&A..52..529Y}}. Thus, we adopt
$\dot{m}=0.1$ as a fiducial value. The accretion rate might be
$\dot{m}>0.1$ during the formation of the FBs. In this case, we assume
that the CRs are injected just before and/or after the formation when
$\dot{m}\sim 0.1$. The mass of the SMBH is $M_{\rm BH}=4.3\times 10^6\:
M_\odot$ \cite{2009ApJ...692.1075G}.

The luminosity of the CR protons accelerated in the RIAF is assumed to
be $L_{\rm p,\rm tot}=\eta_{\rm cr}\dot{M} c^2$, where $\eta_{\rm cr}$
is the parameter governing CR acceleration. We adopt $\eta_{\rm
cr}=2\times 10^{-3}$, motivated by the model for IceCube's neutrinos
\cite{2015ApJ...806..159K}. For the stochastic CR acceleration, the
production rate of CR protons in the momentum range $p$ to $p+dp$ is
written as
\begin{equation}
\label{eq:sp}
 \dot{N}(x)dx \propto x^{(7-3q)/2}K_{(b-1)/2}(x^{2-q})dx\:,
\end{equation}
where $x=p/p_{\rm cut}$, $K_\nu$ is the Bessel function, and $b=3/(2-q)$
\cite{2006ApJ...647..539B}. We assume that the turbulence that is
responsible for the stochastic acceleration is a Kolmogorov type, and
thus the power-law index is $q=5/3$. The cutoff momentum is given by
$p_{\rm cut}=(2-q)^{1/(2-q)}p_{\rm eq}=p_{\rm eq}/27$, where $p_{\rm
eq}=E_{\rm p,eq}/c$ \cite{2015ApJ...806..159K,2006ApJ...647..539B}.

\subsubsection{Diffusion coefficient in the Galactic halo}
\label{sec:diffc}

In this study, we focus on the CRs around the knee and we assume that
they were injected through an explosive activity of Sgr~A*, which
created the FBs about $10$~Myr ago. The diffusion coefficient for the
CRs around the knee energy and its spatial dependence are hardly
known. We consider a two-zone model of the Galaxy, in which the
diffusion coefficient of CRs in the halo $D_{\rm h}$ is different from
that in the disk $D_{\rm d}$
\cite{2012ApJ...752L..13T,2014ApJ...782...34B}. The disk is thin and the
scale height is $H_{\rm d}\sim 0.3$~kpc \cite{2014ApJ...782...34B}.  In
fact, the diffusion coefficient in the disk can be affected by the
turbulence generated by strong magnetic fields, stellar winds, and
supernovae, while their impact is much less in the halo. For example, if
we assume Kolmogorov-type turbulence, their ratio is
\begin{equation}
\label{eq:DD}
 \frac{D_{\rm d}}{D_{\rm h}} 
= \left(\frac{l_{\rm d}}{l_{\rm h}}\right)^{2/3}
\left(\frac{B_{\rm d}}{B_{\rm h}}\right)^{-1/3}
= 0.07\left(\frac{l_{\rm d}/l_{\rm h}}{0.03}\right)^{2/3}
\left(\frac{B_{\rm d}/B_{\rm h}}{3}\right)^{-1/3}\:,
\end{equation}
where $l_{\rm h}$ and $l_{\rm d}$ are the coherent lengths of the halo
($B_{\rm h}$) and the disk magnetic fields ($B_{\rm d}$), respectively
\cite{2012ApJ...757...14J,2016JCAP...05..056B}.

We assume that the Galactic halo is spherically symmetric for the knee
CRs for the sake of simplicity. The outer boundary of the halo, $R_{\rm
h}$, at which CRs escape into the intergalactic space is not known and
it probably depends on the energy of the CRs. Since we assume that the
high-energy CRs are coming from Sgr~A* and their observed arrival
directions on the Earth are almost isotropic
\cite{2013ApJ...765...55A,2017ApJ...836..153A}, the radius of the halo
$R_{\rm h}$ should be larger than the distance to Sgr~A*
($R_\odot=8$~kpc), and thus we assume that $R_{\rm h}\gtrsim 10$~kpc.

Assuming that an explosive ejection of CRs occurred at $t=0$ and the
current time is $t=t_0 (>0)$, we can estimate an appropriate diffusion
coefficient of the CRs in the Galactic halo. The influence of the disk
can be ignored because the disk is thin and the diffusion of high-energy
CRs there is fast enough. In fact, even if the diffusion coefficient for
the disk is much smaller than that for the halo, the ratio of the
diffusion time for the disk, $t_{\rm diff,d}$, to that for the halo near
the Earth, $t_{\rm diff,h\odot}$, is much smaller than one:
\begin{equation}
\frac{t_{\rm diff,d}}{t_{\rm diff,h\odot}}
= \left(\frac{H_{\rm d}^2}{4\: D_{\rm d}}\right)
/\left(\frac{R_\odot^2}{6\: D_{\rm h}}\right)
= 0.02\: \left(\frac{H_{\rm d}}{0.3\rm\: kpc}\right)^2
\left(\frac{R_\odot}{8\:\rm kpc}\right)^{-2}
\left(\frac{D_{\rm d}/D_{\rm h}}{0.1}\right)^{-1}\:,
\end{equation}
where the difference of the numbers in the second expression (1/4 and
1/6) comes from that of the dimension of the disk (2D) and the halo
(3D). The small ratio means that the CR density in the disk and that in
the halo at a given distance from the Galactic center $r$ is almost the
same, because CRs in the disk is almost in equilibrium with those in the
nearby halo. The diffusion coefficient in the halo that gives a
diffusion time $t_0$ and a diffusion scale $R_{\rm h}$ is
\begin{equation}
 D'_{\rm knee} \sim \frac{R_{\rm h}^2}{6\: t_0}
=2.0\times 10^{30}\left(\frac{R_{\rm h}}{\rm 20\: kpc}\right)^2
\left(\frac{t_0}{\rm 10\: Myr}\right)^{-1}\rm cm^2\: s^{-1}\:.
\end{equation}
We adopt a diffusion coefficient that is a few times larger than this so
that a significant fraction of the CRs are allowed to escape from the
halo at $t=10$~Myr. This is because Sgr~A* produces a fairly large
amount of CRs for given parameters (e.g. $\dot{m}$ and $\eta_{\rm
cr}$). Thus, the value we adopt is $D_{\rm knee}=4.4\times 10^{30}\:\rm
cm^2\: s^{-1}$ at the energy of $E=10^{15.5}$~eV. Since $D_{\rm knee}<
c\: l_{\rm h}/3$ is expected, the coherent length may be $l_{\rm
h}>140$~pc. Assuming that the turbulence that scatters CRs is a
Kolmogorov-type, the diffusion coefficient for energies around the knee
is represented by
\begin{equation}
\label{eq:Dh}
 D_{\rm h}(E) = D_{\rm knee}\left(\frac{E}{10^{15.5}\rm eV}\right)^{1/3}
= 3\times 10^{28}\left(\frac{E}{\rm GeV}\right)^{1/3}
\rm cm^2\: s^{-1}\:.
\end{equation}
Note that although we normalized the coefficient at $E=1$~GeV in the
last equation following a convention, it does not mean that the actual
coefficient in the halo is represented by eq.~(\ref{eq:Dh}) down to
$E\sim$~GeV.  If the turbulence is generated by CR streaming
\cite{1975MNRAS.173..255S}, it should reflect the characteristics of the
streaming CRs (e.g. energies of the CRs or spatial distribution of their
sources). Thus, if the GeV CRs are accelerated at supernova remnants
(SNRs) distributed in the Galactic disk for example, the turbulence that
scatters the GeV CRs probably differs from that generated by the knee
CRs accelerated at Sgr~A*. The disk scale $H_{\rm d}$ and the halo scale
$R_{\rm h}$ for the GeV CRs could also be much different from those for
the CRs around the knee.

\subsubsection{Diffusion of CR protons}
\label{sec:diff}

We calculate the diffusion of CR protons in the spherical Galactic
halo. The effects of the Galactic disk can be ignored as we discussed in
the previous subsection. Since energy losses due to hadronic
interactions are negligible for high-energy CRs, the diffusion equation
is
\begin{equation}
\label{eq:diff}
 \frac{\partial f_{\rm B}}{\partial t}
= \frac{1}{r^2}\frac{\partial}{\partial r}
\left(r^2 D_{\rm h}(p)\frac{\partial
					       f_{\rm B}}{\partial r}\right)
\:,
\end{equation}
where $f_{\rm B}=f_{\rm B}(t,r,p)$ is the distribution function for the
CRs injected at the outburst of Sgr~A*. We assume that $f_{\rm B}=0$ at
$r=R_{\rm h}$ and $\partial f_{\rm B}/\partial r=0$ at $r=0$. For the
sake of simplicity, we assume that the CRs are instantaneously injected
at the galactic center at $t=0$, although we put them in a tiny sphere
($r<r_{\rm s}\ll R_\odot$) in actual calculations. Since the total
amount of the CR protons accelerated in Sgr~A* is $L_{\rm p, tot} t_{\rm
inj}$, where $t_{\rm inj} (\ll t_0)$ is the duration of the intensive
activity of Sgr~A* that gave birth to the FBs, the distribution function
at $t=0$ is determined from the relation of
\begin{equation}
\label{eq:fint}
 4\pi p^3 c f_{\rm B}(0,r,p)dp = 
\frac{3 L_{\rm p, tot}t_{\rm inj}}{4\pi r_{\rm s}^3}
x\dot{N}(x)dx/\left(\int_0^\infty x\dot{N}(x)dx \right)\:,
\end{equation}
for $r<r_{\rm s}$ and $f_{\rm B}(0,r,p) = 0$ for $r>r_{\rm s}$. We
analytically solve eq.~(\ref{eq:diff}) and the derivation of the
solutions are shown in the Appendix. We assume $t_{\rm inj}=1$~Myr
because it makes $\dot{m}\sim 0.1$ compatible with appropriate proton
production. We also assume that $r_{\rm s}=0.1$~kpc, although the value
is not important as long as we consider $r\gg r_{\rm s}$ for the halo
CRs. Since $L_{\rm p, tot}t_{\rm inj}\propto \eta_{\rm cr}\:\dot{m}\:
t_{\rm inj}$, combinations of $\eta_{\rm cr}$, $\dot{m}$, and $t_{\rm
inj}$ that give the same $L_{\rm p, tot}t_{\rm inj}$ give the same
results.

We ignore the advection of CRs by galactic winds. In fact, using
eq.~(\ref{eq:Dh}), the ratio of the advection time ($t_{\rm adv}$) to
the diffusion time in the halo ($t_{\rm diff,h}$) is represented by
\begin{eqnarray}
\frac{t_{\rm adv}}{t_{\rm diff,h}}&=&
 \left(\frac{R_h}{V_{\rm w}}\right)
/\left(\frac{R_{\rm h}^2}{6\: D_{\rm h}(E)}\right)\nonumber\\
&\sim& 9\left(\frac{V_{\rm w}}{500\rm\: km\: s^{-1}}\right)^{-1}
\left(\frac{R_{\rm h}}{20\rm\: kpc}\right)^{-1}
\left(\frac{E}{10^{15.5}\rm\: eV}\right)^{1/3}\:,
\end{eqnarray}
where $V_{\rm w}$ is the wind velocity. Since the ratio is larger than
one for wind velocities for normal star-forming galaxies ($V_{\rm
w}\lesssim 500\rm\: km\: s^{-1}$
\cite{2006ApJ...646..951K,2015ApJ...806...24S}), the advection of CRs by
the galactic winds is not important for $E\gtrsim$~TeV.

\begin{figure}[tbp]
\centering \includegraphics[width=90mm]{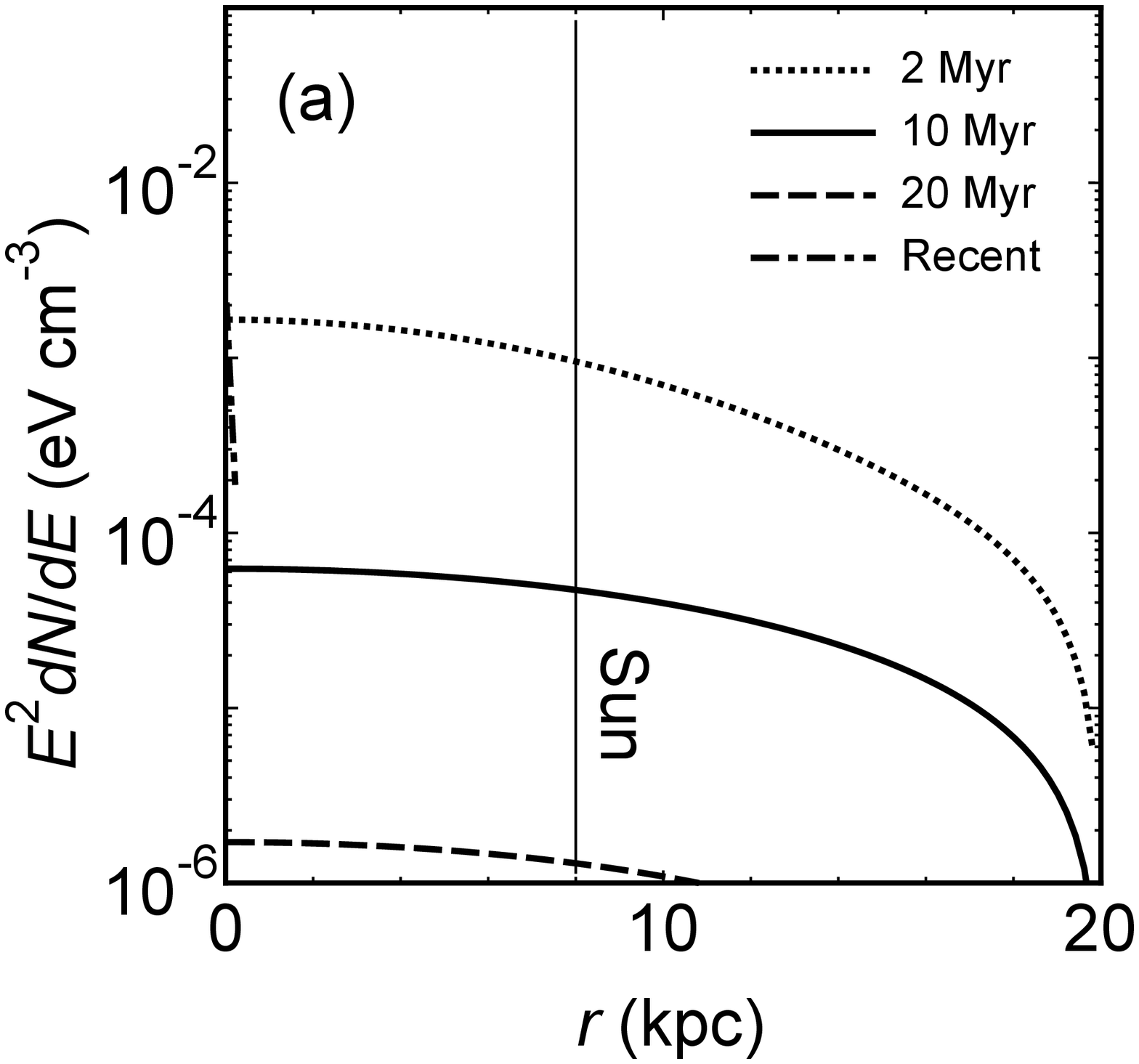}
\includegraphics[width=90mm]{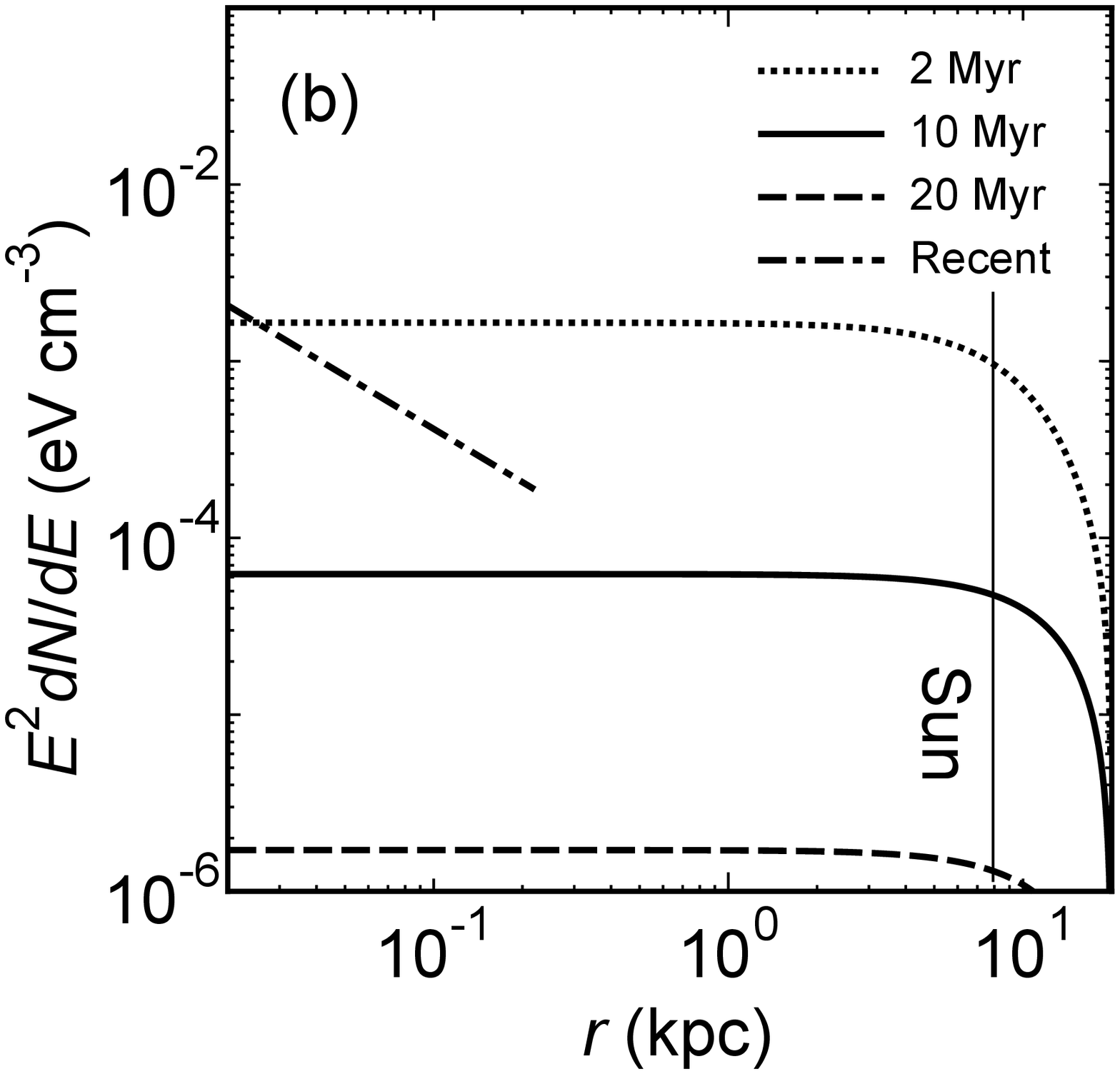} \caption{\label{fig:rad} Radial
profiles of CR density at $E=10^{15.5}$~eV for $t=2$~Myr (dotted),
10~Myr (thick-solid), and 20~Myr (dashed). (a) linear plot, and (b)
logarithmic plot. The contribution of CRs that recently enter into the
CMZ is shown by the dot-dashed line (see section~\ref{sec:CMZ}). The
position of the solar system is shown by the vertical thin-solid lines.}
\end{figure}

\subsection{CRs from supernova remnants}

It is likely that CRs with lower energies ($\lesssim 100$~TeV) come from
SNRs. For the demonstrative purpose, we consider their contribution
assuming that the SNRs are steady CR sources. The CR injection rate is
represented by
\begin{eqnarray}
\label{eq:QSNR}
Q_{\rm SNR}(p, R, z) 
&=& Q_0 \left(\frac{p}{p_0}\right)^{-\mu}
e^{-\frac{p}{p_{\rm SNR}}}
\frac{\beta^2 e^{-\beta}}{12\pi R_\odot}
\left(\frac{R}{R_\odot}\right)^2 \nonumber \\
&\times& \exp\left[-\frac{\beta(R-R_\odot)}{R_\odot}\right]\frac{1}{h}
e^{-\frac{|z|}{h}}\:,
\end{eqnarray}
where $Q_0=5\times 10^{13}{\rm\: s^{-1}\: kpc^{-1}}\: ({\rm
GeV}/c)^{-3}$, $\mu=4.35$, $p_0 c = 3\times 10^{15}$~eV, $p_{\rm
SNR} c = 1\times 10^{15}$~eV, $\beta=3.53$, and $h=0.1$~kpc
\cite{1996A&AS..120C.437C,2012JCAP...01..010B}. We treat the diffusion
of the CRs from SNRs separately from that for the CRs from Sgr~A* for
the sake of simplicity, and we take the cylindrical spatial coordinates
represented by $R$ and $z$, where $R$ is the distance from the symmetry
axis and $z=0$ is the symmetry plane of the Galactic disk. The boundary
of the Galactic halo is set at $R=R_{\rm d}=20$~kpc and $z=H_{\rm
h}(E)$. Since CRs with higher energies tend to create
magnetic fluctuations that resonate with them on a wider scale
(e.g. Ref.~\cite{2011MNRAS.415.3434F}), the halo height could be
modeled as
\begin{equation}
\label{eq:HhE}
H_{\rm h}(E) = \left\{\begin{array}{ll}
		H_{\rm h1} &\mbox{for $E<$~TeV}\\
		   \frac{H_{\rm h2}- H_{\rm h2}}{\log(1000)}
		    \log(E/{\rm TeV}) + H_{\rm h1}
		    &\mbox{for TeV$\leq E\leq$~PeV}\\
		H_{\rm h2} &\mbox{for $E>$~PeV}
		     \end{array}\right.
\:,
\end{equation}
where $H_{\rm h1}=10$~kpc and $H_{\rm h2}=20$~kpc for the fiducial
model.  We chose $p_{\rm SNR}$ so that the CR spectrum observed on the
Earth is reproduced. The diffusion coefficient $D_{\rm SNR}$ for
$|z|>H_{\rm d}=0.3$~kpc is given by eq.~(\ref{eq:Dh}) and that for
$|z|<H_{\rm d}$ is given by $D_{\rm d}$, which is specified later
(eq.~\ref{eq:Dd}). The diffusion equation is
\begin{equation}
 \label{eq:diff_snr}
 \frac{\partial f_{\rm SNR}}{\partial t}
= \frac{1}{R}\frac{\partial}{\partial R}
\left(D_{\rm SNR} R\frac{\partial f_{\rm SNR}}{\partial R}\right)
+ \frac{\partial}{\partial z}\left(D_{\rm SNR}
\frac{\partial f_{\rm SNR}}{\partial z}\right)+ Q_{\rm SNR}\:,
\end{equation}
where $f_{\rm SNR}=f_{\rm SNR}(t,r,p)$ is the distribution function for
the SNR component. Since we consider a steady state, the left-hand side
of eq.~(\ref{eq:diff_snr}) is zero.

\section{Results for the fiducial model}
\label{sec:result}

In this section, we show the results of our fiducial model that adopts
the parameters we explained in the previous sections.
Figure~\ref{fig:rad} shows the radial density profiles of CRs originated
from the outburst of Sgr~A*. The lines are for $E=10^{15.5}$~eV at
$t=2$, 10 and 20~Myr. The boundary of the Galactic halo is located at
\begin{equation}
\label{eq:RhE}
R_{\rm h}(E) = \left\{\begin{array}{ll}
		R_{\rm h1} &\mbox{for $E<$~TeV}\\
		   \frac{R_{\rm h2}- R_{\rm h2}}{\log(1000)}
		    \log(E/{\rm TeV}) + R_{\rm h1}
		    &\mbox{for TeV$\leq E\leq$~PeV}\\
		R_{\rm h2} &\mbox{for $E>$~PeV}
		     \end{array}\right.
\:,
\end{equation}
where $R_{\rm h1}=10$~kpc and $R_{\rm h2}=20$~kpc for the fiducial
model. At $t\gtrsim 10$~Myr, the radial gradient of the density has
become small; the ratio of the density at the Galactic center ($r=0$) to
that at the Earth ($r=R_\odot=8$~kpc) is only 1.3 at $t=10$~Myr. For
comparison, we showed the density profile of the recently injected
component (see section~\ref{sec:CMZ}). At the Galactic center, this
component is dominant for $t\gtrsim 2$~Myr.

\begin{figure}[tbp]
\centering \includegraphics[width=90mm]{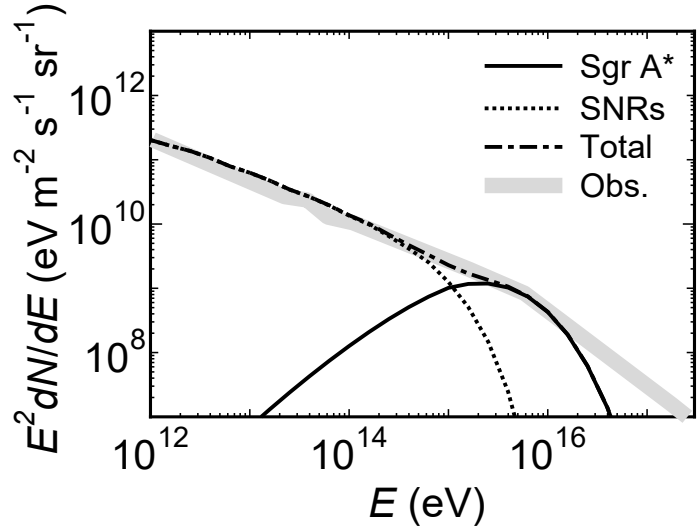} \caption{\label{fig:sp}
The spectrum of CRs injected by Sgr~A* at $r=R_\odot$ at $t=10$~Myr
(solid line). The contribution of CRs from SNRs is shown by the
dotted-line and the total spectrum is shown by the dot-dashed
line. The size of the halo is given by eqs.~(\ref{eq:HhE}) and
(\ref{eq:RhE}). Observations are shown by the gray band
\cite{2011ARA&A..49..119K}.}
\end{figure}

\begin{figure}[tbp]
\centering \includegraphics[width=90mm]{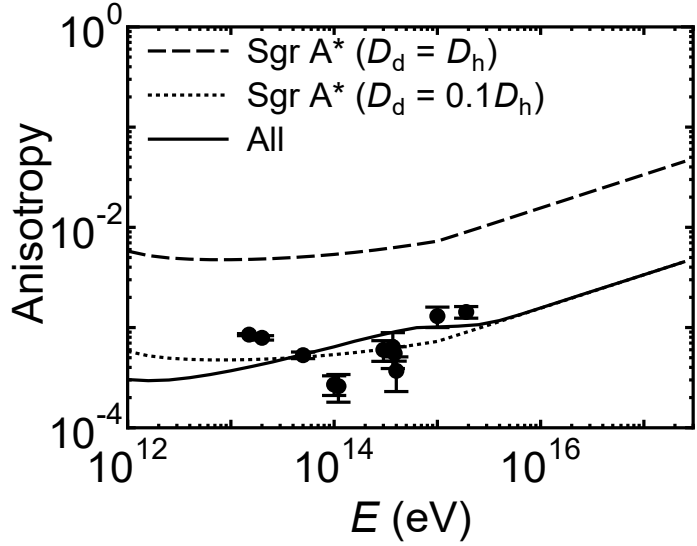}
\caption{\label{fig:aniso} Dipole anisotropy of the arrival directions
of CRs on the Earth ($r=R=R_\odot$, $z=0$). The dashed and the dotted
lines are for the CRs injected by Sgr~A* at the outburst 10~Myrs ago
($f=f_{\rm B}$; $t=10$~Myr). The dashed line is the result when $D_{\rm
d}=D_{\rm h}$ and the dotted line is the one when $D_{\rm d}=0.1\:
D_{\rm h}$. The solid line is for all the components ($f=f_{\rm
B}+f_{\rm SNR}$) and $D_{\rm d}=0.1\: D_{\rm h}$. Recent observations
with EAS-TOP \cite{2009ApJ...692L.130A}, IceCube
\cite{2012ApJ...746...33A}, IceTop \cite{2013ApJ...765...55A}, and Tibet
\cite{2017ApJ...836..153A} are shown by the black dots.}
\end{figure}

Figure~\ref{fig:sp} shows the CR spectrum around the Earth
($r=R=R_\odot$, $z=0$) at $t=10$~Myr. The CRs injected by Sgr~A* at the
outburst ($f_{\rm B}$) contribute to the spectrum around the knee. The
lower-energy CRs are provided by SNRs in the Galactic disk ($f_{\rm
SNR}$). The CRs for $E\gtrsim 10^{16}$~eV may be provided by
extragalactic AGNs or some sources in the Galaxy (see
section~\ref{sec:high}). Figure~\ref{fig:aniso} shows the dipole
anisotropy of the arrival directions of CRs on the Earth ($r=R_\odot$),
which is given by
\begin{equation}
\label{eq:aniso}
a = \frac{3\: D_{\rm d}}{c}\frac{\nabla f}{f}\:,
\end{equation}
where $f$ is the distribution function in general, although strictly
speaking, the anisotropy in eq.~(\ref{eq:aniso}) is not the same as the
observed projected anisotropy. The dashed line is the anisotropy for the
outburst component of Sgr~A* ($f=f_{\rm B}$) and $D_{\rm d}=D_{\rm h}$,
where $D_{\rm h}$ is given by eq.~(\ref{eq:Dh}). The anisotropy around
the knee ($E\sim 10^{15.5}$~eV) is several times larger than the
observations (dashed line). The weak bend at $E\sim 10^{14}$~eV is made
because the diffusion radius ($\sim \sqrt{ D_{\rm h}(E)t}$) is
comparable to $R_\odot$ around this energy. The discrepancy between the
model and the observations can readily be solved if $D_{\rm d}$ is
several times smaller than $D_{\rm h}$, as is theoretically expected
(eq.~\ref{eq:DD}). If we shift the dashed line vertically, we can
approximate the anisotropy when $D_{\rm d}\neq D_{\rm h}$. For example,
if we shift the dashed line by multiplying 0.1, the shifted line
approximates the anisotropy when $D_{\rm d}=0.1\: D_{\rm h}$ or
\begin{equation}
\label{eq:Dd}
 D_{\rm d}(E) 
= 3\times 10^{27}\left(\frac{E}{\rm GeV}\right)^{1/3}
\rm cm^2\: s^{-1}\:.
\end{equation}
The prediction is close to the observations (dotted-line). Here, we
assumed that the CR density in the the disk is the same as that outside
the disk at a given $r$ (section~\ref{sec:diffc}). This type of a thin
disk with a smaller diffusion coefficient has been considered in
previous studies and is consistent with observations of the ratio
between secondary and primary particles with lower energies ($E\lesssim
10^{12}$~eV) if the effects of Galactic spiral arms are considered
\cite{2014ApJ...782...34B}, although observations have not well
constrained $H_{\rm d}$ and $D_{\rm d}$ for CRs around the knee. The
solid line shows the CR anisotropy for all the components ($f=f_{\rm
B}+f_{\rm SNR}$). The SNR component ($f_{\rm SNR}$) is dominant for
$E\lesssim 10^{14.5}$~eV (figure~\ref{fig:sp}). The anisotropy of
the SNR component is not much different from that of the outburst
component. The CRs from SNRs are injected by multiple sources, which
reduces the anisotropy, while their local sources are distributed along
the Galactic plane, which increases the anisotropy. The CRs from Sgr~A*
is injected by a single source, which increases the anisotropy, while
the source is distant and the CRs from it are filled in the large halo
($R_{\rm h}\sim 10$--20~kpc), which reduces the anisotropy.

\begin{figure}[tbp]
\centering \includegraphics[width=90mm]{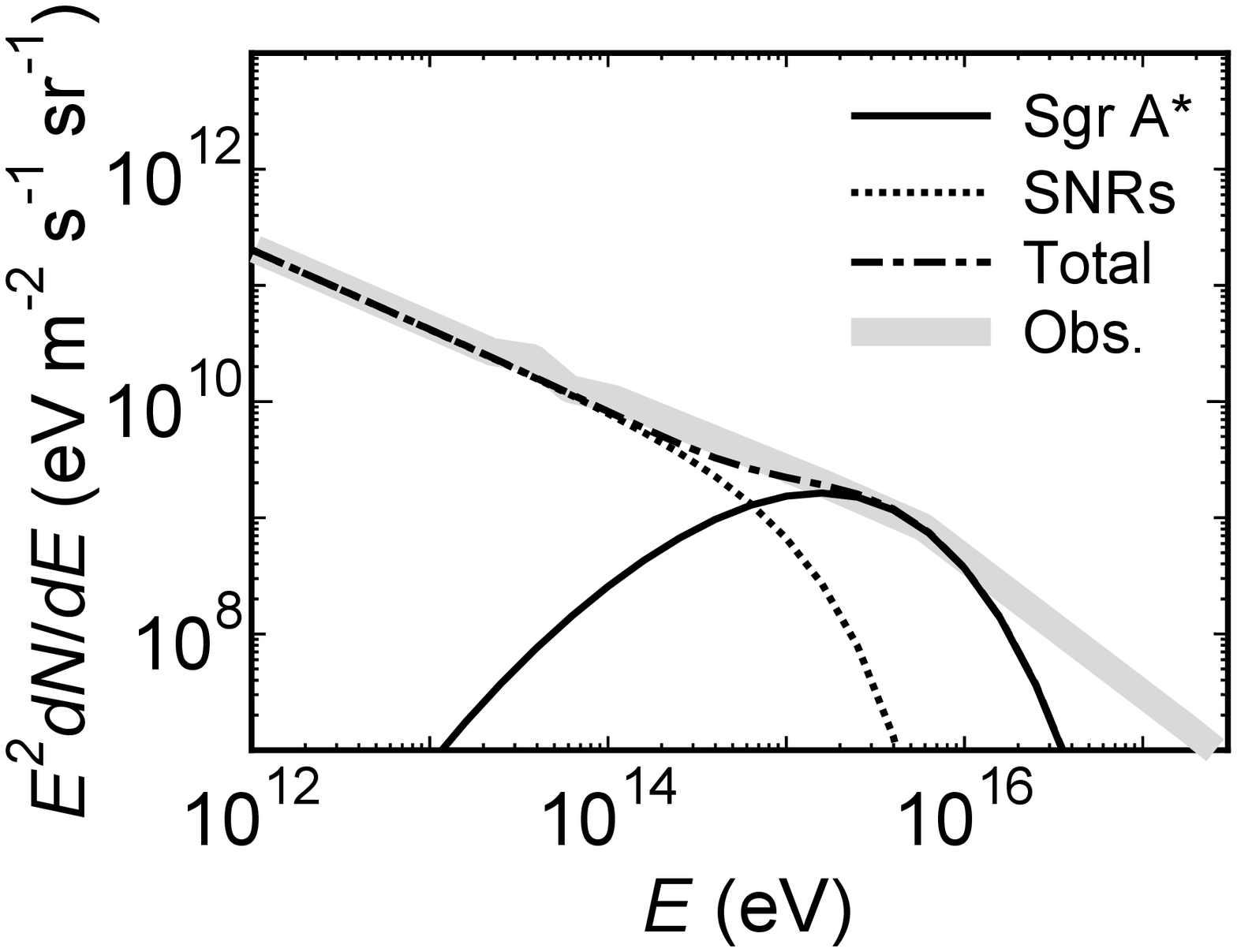}
\caption{\label{fig:sp2} Same as figure~\ref{fig:sp} but for $R_{\rm
h}=10$~kpc and $t=3$~Myr.}
\end{figure}

\begin{figure}[tbp]
\centering \includegraphics[width=90mm]{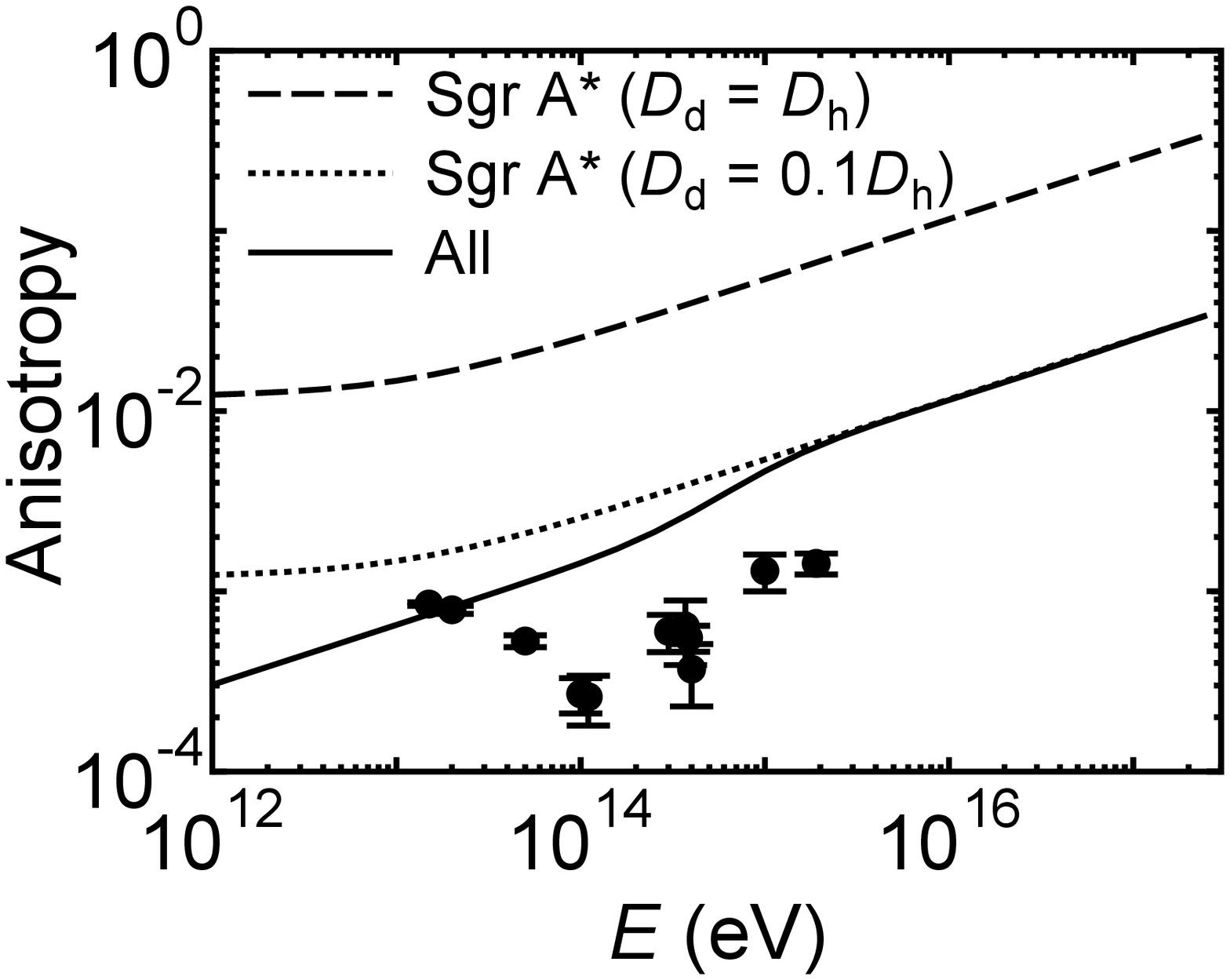}
\caption{\label{fig:aniso2} Same as figure~\ref{fig:aniso} but for
$R_{\rm h}=10$~kpc and $t=3$~Myr.}
\end{figure}

\begin{figure}[tbp]
\centering \includegraphics[width=90mm]{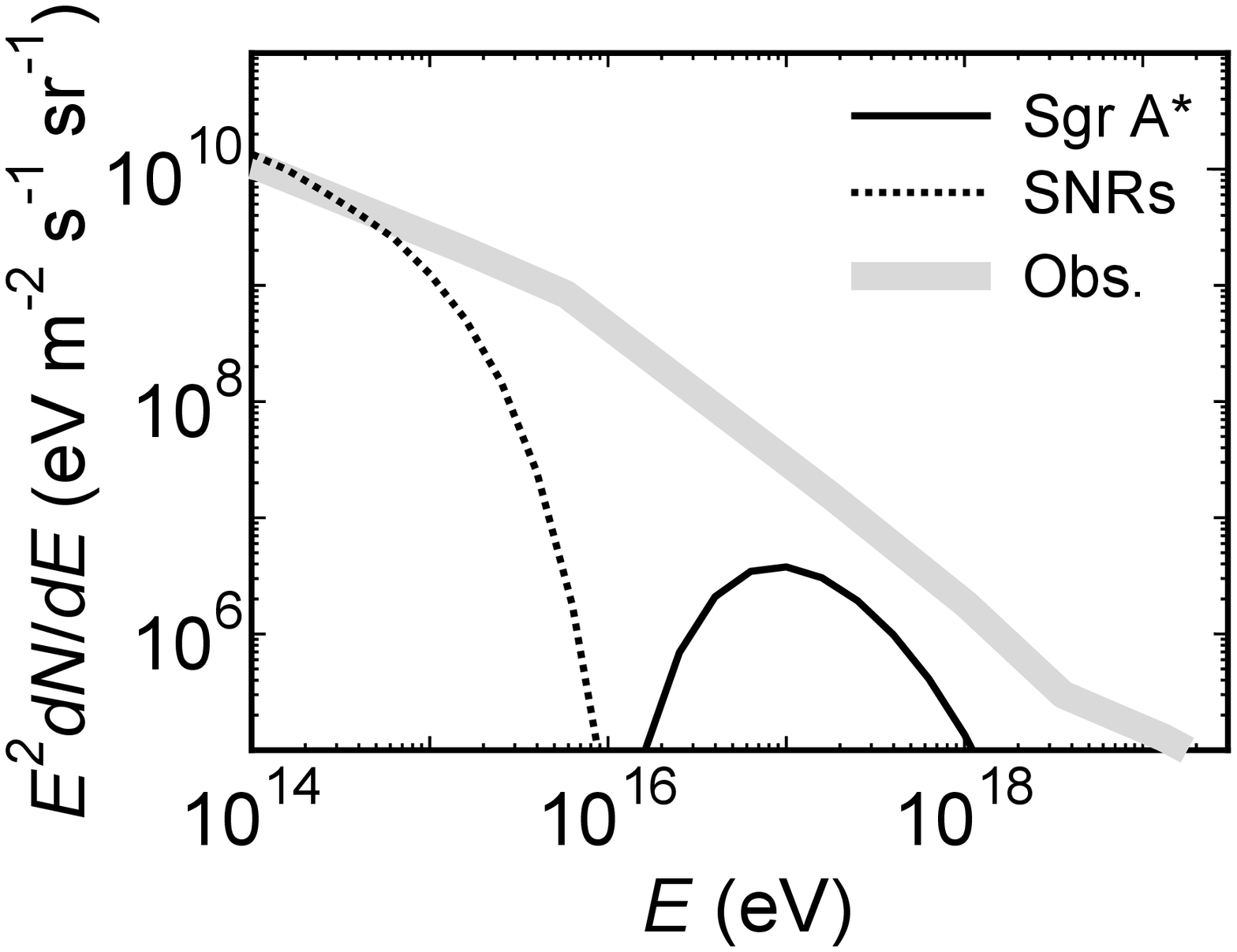}
\caption{\label{fig:spp} Same as figure~\ref{fig:sp} but for the
power-law injection (eq.~\ref{eq:spp}) at higher energies and $t=3$~Myr.
The line for the total spectrum is omitted.}
\end{figure}

\section{Discussion}
\label{sec:disc}

\subsection{B/C ratio}

If the origin of CRs around the knee is different from that of CRs with
lower energies (e.g. SNRs in the Galactic disk), the ratio of secondary
to primary CR abundances can also be different. Since our model is a
single-source, single-burst scenario, we can predict the ratio fairly
easily.  Here, we focus on the B/C ratio. Assuming that the
Galactic halo is represented by an one-zone model (leaky-box-like)
and that the influence of the disk can be ignored, the evolution of the
total number of boron $N_{\rm B}(t)$ in the halo is written as
\begin{equation}
\label{eq:NB}
 \frac{\partial N_{\rm B}}{\partial t} = 
- \frac{N_{\rm B}}{t_{\rm esc}}
- \frac{N_{\rm B}}{\tau_{\rm B}}
+ \frac{N_{\rm C}}{\tau_{\rightarrow \rm B}}\:,
\end{equation}
where $N_{\rm C}(t)$ is the carbon number, $t_{\rm esc}=R_{\rm
h}^2/(6\rm D_{\rm h})$ is the escape time-scale of CRs from the halo,
$\tau_{\rm B}$ is the spallation time-scale for boron, and
$\tau_{\rightarrow \rm B}$ is the effective production time-scale for
boron. We assume that primary CRs are instantaneously injected at $t=0$.
Since the fraction of carbon that changes into other elements is
ignorable for our fiducial model, the number of carbon evolves as
$N_{\rm C}\propto \exp(-t/t_{\rm esc})$. Thus, eq.~(\ref{eq:NB}) gives
the B/C ratio of
\begin{equation}
 R_{\rm B/C}(t) \equiv\frac{N_{\rm B}(t)}{N_{\rm C}(t)} = 
\frac{\tau_{\rm B}}{\tau_{\rightarrow \rm B}}(1-e^{-t/\tau_{\rm B}})\:,
\end{equation}
assuming that $N_{\rm B}=0$ at $t=0$. The leaky-box-like model
represented by eq.~(\ref{eq:NB}) would be a reasonably good
approximation of the diffusion model for $t\gtrsim t_{\rm esc}$,
because the halo is filled with the CRs. The time-scales are related to
grammages such as $\Lambda_{\rm B}=\beta c\rho\tau_{\rm B}$ and
$\lambda_{\rightarrow \rm B} = \beta c\rho\tau_{\rightarrow \rm B}$,
where $\beta = v/c~ (\sim 1)$ is the particle velocity normalized by the
light velocity, and $\rho$ is the typical gas density in the Galactic
halo. Recent X-ray observations showed that $\rho\sim 0.002\: m_{\rm
p}\: \rm cm^{-3}$ at $r\sim R_\odot/2=4$~kpc \cite{2015ApJ...807...77K}.
The escape time-scale is $t_{\rm esc}=4.6$~Myr for $R_{\rm h}=20$~kpc
and $E=10^{15.5}$~eV (eq.~\ref{eq:Dh}). If we use $\Lambda_{\rm
B}=9.3\:\rm g\: cm^{-2}$ and $\lambda_{\rightarrow \rm B}=26.8\:\rm g\:
cm^{-2}$ \cite{1990PhRvC..41..533W,2012ApJ...752...69O}, the B/C ratio
is $R_{\rm B/C}(10\rm\: Myr)\sim 0.001$, and it is independent of $E$.
Observed B/C ratios can be approximated as $R_{\rm B/C,ext}\sim 0.2\:
(E/{10\rm\: GeV})^{-1/3}$ for $10\lesssim E\lesssim 1000$~GeV
\cite{2012ApJ...752...69O,2014ApJ...782...34B}. If the B/C ratio is
extrapolated to $E\sim 10^{15.5}$~eV, it is $R_{\rm B/C,ext}\sim 0.003$
and is comparable to the predicted $R_{\rm B/C}$. If the observed B/C
ratio is extrapolated as $R_{\rm B/C,ext}\sim 0.3\: (E/{10\rm\:
GeV})^{-0.6}$ for $E\gtrsim 1$~TeV as some models suggest
\cite{2012ApJ...752...69O,2014ApJ...782...34B}, the ratio is $R_{\rm
B/C,ext}\sim 10^{-4}$ at $E\sim 10^{15.5}$~eV, which is much smaller
than the predicted $R_{\rm B/C}$. Anyway, a bend would be observed in
the energy-B/C ratio relation as $E$ approaches the knee from smaller
energies because the CRs from Sgr~A* become dominant
(figure~\ref{fig:sp}).

\subsection{Halo size}

Since little is known about the size of the Galactic halo that confines
the CRs around the knee, we study the CR diffusion when $R_{\rm h}$ is
changed. Figure~\ref{fig:sp2} is the CR spectrum at $r=R_\odot$ when
$R_{\rm h}=R_{\rm h2}=R_{\rm h1}=10$~kpc. Since CRs escape more
easily and the CR density decreases faster than when $R_{\rm
h2}=20$~kpc, we chose $t=3$~Myr. The spectrum is not much different
from that in figure~\ref{fig:sp}. However, figure~\ref{fig:aniso2} shows
that the anisotropy becomes larger than that in
figure~\ref{fig:aniso}. Thus, models with a smaller $R_{\rm h}$ require
a smaller $D_{\rm d}$ as well as a younger age of the FBs. Note that the
flux of the CRs from SNRs is slightly smaller than that in
figure~\ref{fig:sp} because those CRs also escape from the halo more
easily.

\subsection{Higher energy component}
\label{sec:high}

Recent observations have shown that CRs with higher energies ($E\sim
10^{17}$--$10^{17.5}$~eV) contain a significant fraction ($\sim 80$~\%)
of light elements (H and He), which suggests that they have a Galactic
origin \cite{2016Natur.531...70B}. We briefly discuss whether those CRs
are provided through outbursts of Sgr~A*. We use eq.~(\ref{eq:Dh}) as
the diffusion coefficient for the halo. However, since $D_{\rm h}$ is
the increasing function of CR energy, CRs with energies much larger than
the knee ($E\sim 10^{17}$~eV) have already left the Galactic halo when
the knee CRs arrive on the Earth ($t_0\sim 10$~Myr). Thus, it is
difficult to attribute both the CRs around the knee and those above the
knee to the same outburst about 10~Myr ago, unless an extremely large
amount of CRs above the knee are generated at that outburst. Thus, we
assume that the CRs with $E\sim 10^{17}$~eV were injected at another
more recent outburst of Sgr~A*.

Here, we do not confine acceleration mechanisms to that we explained in
section~\ref{sec:CRacc}. Thus, we adopt a generalized power-law spectrum
for the CRs accelerated at Sgr~A*, %which is given by eq.~(\ref{eq:spp}).
\begin{equation}
\label{eq:spp}
 \dot{N}(p)\propto \left(\frac{p}{p_{\rm S}}\right)^{2-\mu} \exp
\left(-\frac{p_{\rm S}}{p}-\frac{p}{p_{\rm L}}\right)\:.
\end{equation}
We assume $\mu=4$ and we set the lower and upper cutoff momenta at
$p_{\rm S}c=10^{17}$~eV and at the 'ankle' ($p_{\rm L}c=10^{18.5}$~eV),
respectively. The initial distribution function for $r<r_{\rm s}$ is
given by eq.~(\ref{eq:fint}),
%
%\begin{equation}
%\label{eq:fint2}
% f(0,r,p) = \frac{3 L_{\rm p, tot}t_{\rm inj}}{4\pi r_{\rm s}^3}
%\frac{p}{p_{\rm S}^2} \dot{N}\:,
%\end{equation}
%
and it is zero for $r>r_{\rm s}$. We assume that $x=p/p_{\rm S}$,
$t_{\rm inj}=1$~Myr, and $r_{\rm s}=0.1$~kpc. We adopt $\dot{m}=3\times
10^{-4}$ and $\eta_{\rm cr}=2\times 10^{-3}$. Figure~\ref{fig:spp} shows
the CR spectrum on the Earth at $t=3$~Myr. Since $L_{\rm p, tot}t_{\rm
inj}$ is 0.3\% of that for figure~\ref{fig:sp} ($\dot{m}=0.1$,
$\eta_{\rm cr}=2\times 10^{-3}$, and $t_{\rm inj}=1$~Myr), the scale of
this outburst is relatively small. Although the injected spectrum is a
power-law (eq.~\ref{eq:spp}), the spectrum on the Earth is not. This is
because part of higher-energy CRs ($E\sim 10^{18}$~eV) have already
escaped from the halo. On the other hand, the observed CR spectrum
between the knee and the ankle is represented by a power-law
(figure~\ref{fig:spp}). Thus, the CR spectrum of this energy range may
be superposed by CRs produced through multiple small outbursts of Sgr~A*
with various $\mu$, $p_{\rm S}$, and $p_{\rm L}$. Although the predicted
spectrum is smaller than the observation in figure~\ref{fig:spp}, the
former can be more close to the latter if we adopt a larger $\dot{m}$,
$\eta_{\rm cr}$, or $t_{\rm inj}$. However, heavy elements that have a
different origin may partially contribute to the observed spectrum
\cite{2013PhRvD..87h1101A,2016A&A...595A..33T}.

\begin{figure}[tbp]
\centering \includegraphics[width=90mm]{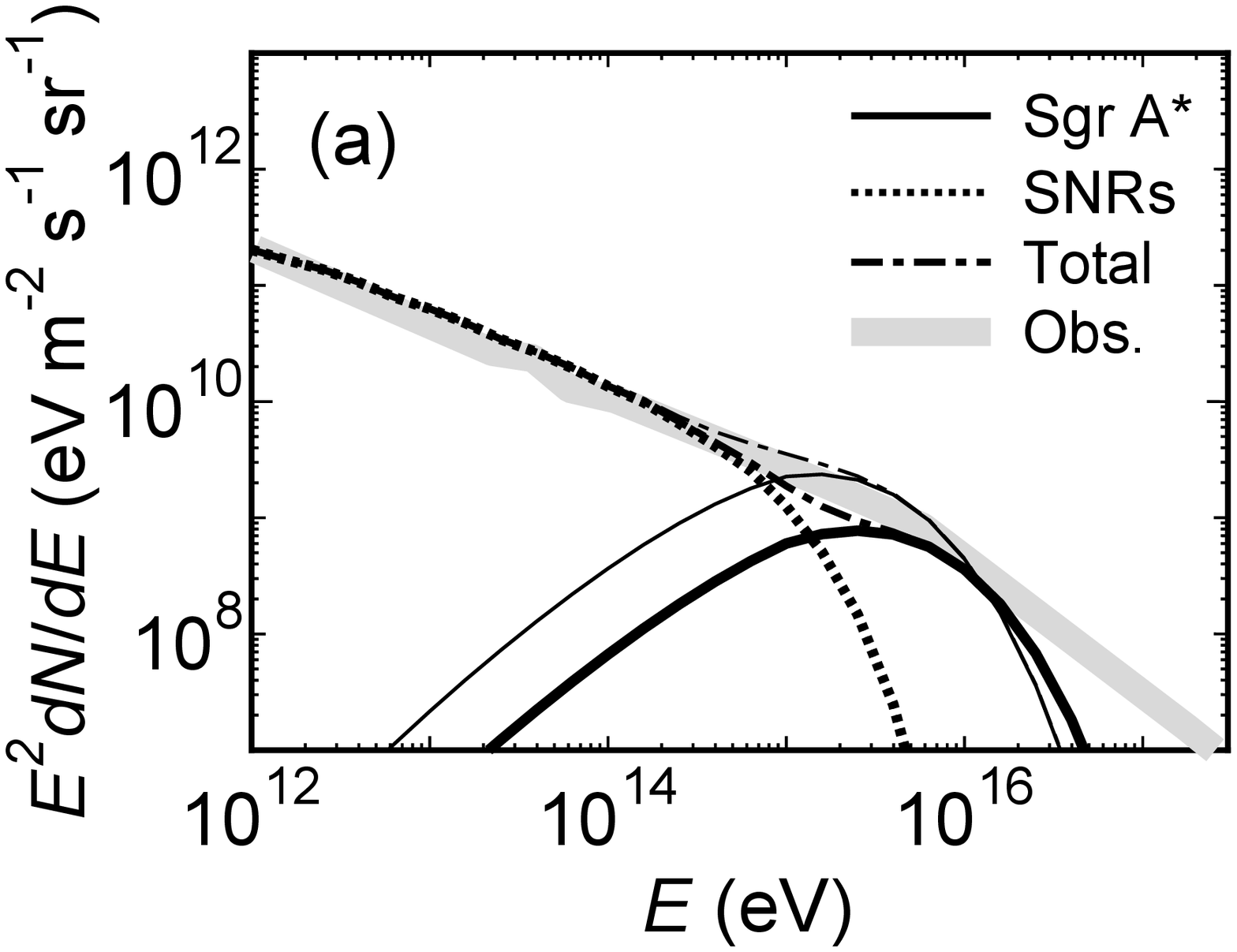}
\includegraphics[width=90mm]{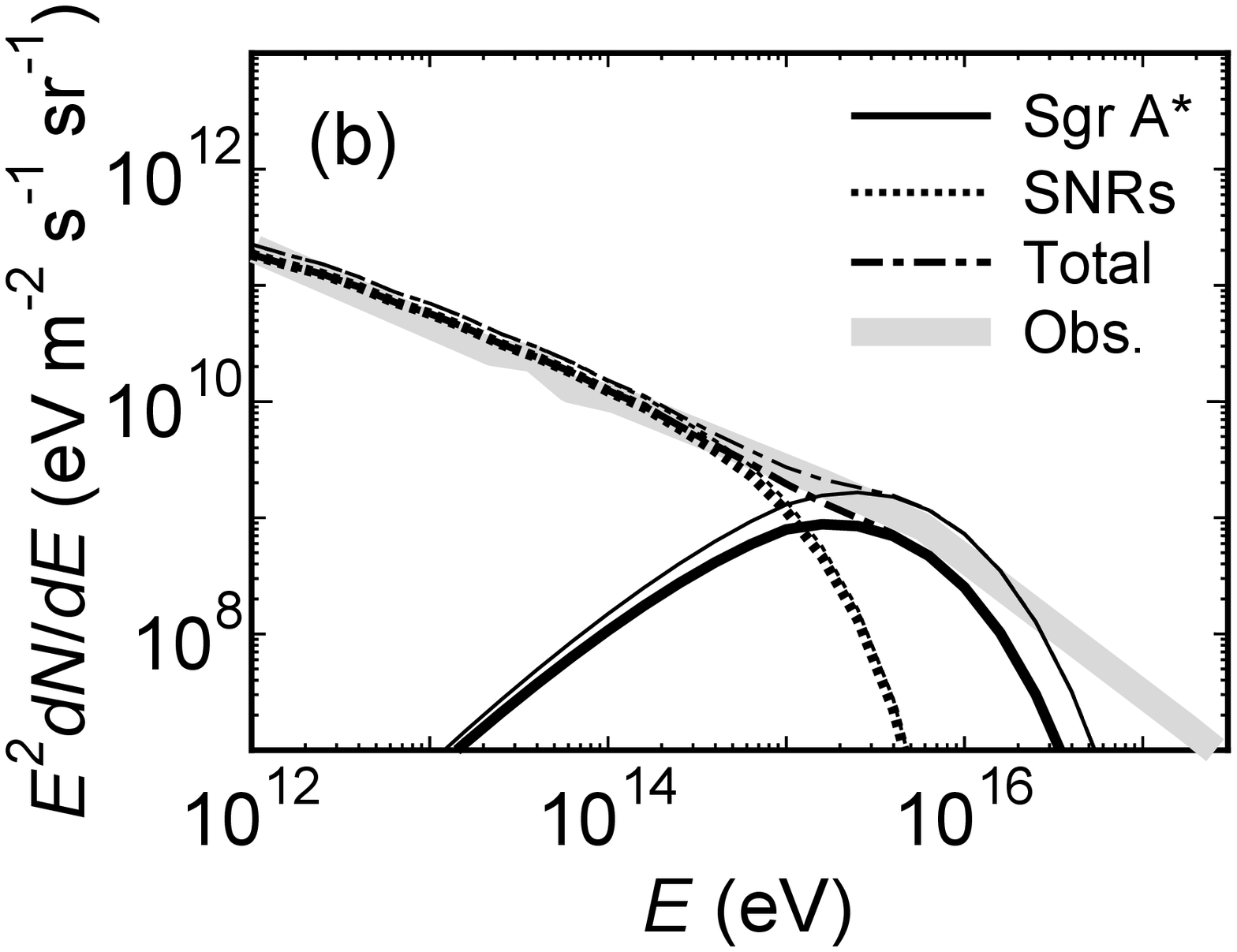} \caption{\label{fig:allow} Same as
figure~\ref{fig:sp} but parameters are slightly changed.  (a) $p_{\rm
eq}=1.5\: p_{\rm eq,fid}$ (thick) and $p_{\rm eq}=0.5\: p_{\rm eq,fid}$
(thin). (b) $D_{\rm h}=1.1\: D_{\rm h,fid}$ (thick) and $D_{\rm h}=0.9\:
D_{\rm h,fid}$ (thin).}
\end{figure}

\subsection{Allowed parameter regions}

In this study, we have assumed that the CR spectrum on the Earth
has three components: the SNR component ($E\lesssim 10^{14.5}$~eV), the
outburst component ($10^{14.5}\lesssim E\lesssim 10^{16}$~eV), and the
possible high energy component ($E\gtrsim 10^{16}$~eV). However, the
observed spectrum is represented by a smooth power-law with a relatively
sharp break at the knee ($E\sim 10^{15.5}$~eV) as if the spectrum were
composed of only two components ($E\lesssim 10^{15.5}$~eV and $E\gtrsim
10^{15.5}$~eV). This means that the parameters related to the outburst
component (the typical energy of CRs, the CR luminosity, and the
diffusion coefficient in the halo) must be fine-tuned.

Figure~\ref{fig:allow}a shows the CR spectrum when the typical energy of
CRs is given by $p_{\rm eq}=1.5\: p_{\rm eq,fid}$ or $0.5\: p_{\rm
eq,fid}$, where $p_{\rm eq,fid}$ is the fiducial value adopted in
figure~\ref{fig:sp} (eq.~\ref{eq:geq}). Other parameters are the same as
those for figure~\ref{fig:sp}. If we change $p_{\rm eq}$ further while
fixing the SNR component, the sharp break at the knee is impaired. Thus,
the spectrum would look like being composed of three components if we
include a fixed high energy component ($E\gtrsim 10^{16}$~eV). The same
happens when the coefficient of the CR luminosity is changed to be
$\eta_{\rm cr}\gtrsim 1.5\: \eta_{\rm cr,fid}$ or $\eta_{\rm cr}\lesssim
(2/3)\: \eta_{\rm cr,fid}$, where $\eta_{\rm cr,fid}=2\times 10^{-3}$ is
the fiducial value adopted in
figure~\ref{fig:sp}. Figure~\ref{fig:allow}b shows the CR spectrum when
the diffusion coefficient in the halo is given by $D_{\rm h}=1.1\:
D_{\rm h,fid}$ or $0.9\: D_{\rm h,fid}$, where $D_{\rm h,fid}$ is the
fiducial value adopted in figure~\ref{fig:sp} (eq.~\ref{eq:Dh}). If we
further change the diffusion coefficient, the spectrum shifts
notably. In this case, however, the overall shape of the spectrum does
not much change and the sharp break at the knee is conserved. 

\subsection{Gamma-rays from other galaxies}

We have shown that a past explosive activity of Sgr~A* can provide CRs
around the knee observed on the Earth. Since those CRs have been
distributed widely in the Galaxy and they are not strongly concentrated
around Sgr~A* (figure~\ref{fig:rad}), it would be difficult to confirm our
model based on their distribution. However, similar AGN activities may
be happening in other galaxies. If we observe those galaxies during or
just after the outburst of the nucleus, ejected PeV CRs may be
concentrated around their center. Moreover, if those galaxies have
enough amount of molecular gas around their center and if it serves as
the target of $pp$-interaction, gamma-rays of sub-PeV could be
produced. Although most of them are absorbed by the extragalactic
background light, photons of a few tens of TeV produced by
electromagnetic cascades could be observable in the future as follows.

The total energy of protons injected by an AGN is written as
\begin{eqnarray}
\label{eq:Lptot}
 L_{\rm p,tot}t_{\rm inj}
&=& \eta_{\rm cr} \dot{m} L_{\rm Edd} t_{\rm inj}\nonumber \\
&=&3.5\times 10^{54}
\left(\frac{\eta_{\rm cr}}{2\times 10^{-3}}\right)
\left(\frac{\dot{m}}{0.1}\right)
\left(\frac{L_{\rm Edd}}{5.5\times 10^{44}\rm\: erg\: s^{-1}}\right)
\left(\frac{t_{\rm inj}}{\rm Myr}\right)
\rm\: erg \:.
\end{eqnarray}
We normalized the Eddington luminosity by that for Sgr~A*. The time
scale of $pp$-interaction is given by
\begin{eqnarray}
 t_{pp}&\sim& 1/(\sigma_{pp}n_{\rm MC}c)\nonumber \\
&=& 1.6\times 10^5 
\left(\frac{n_{\rm MC}}{100\rm\: cm^{-3}}\right)^{-1}\rm\: yr\:,
\end{eqnarray}
where $\sigma_{pp}$ is the inelastic cross section for the
$pp$-interaction, and $n_{\rm MC}$ is the number density of target
protons in the molecular gas. We assumed that $\sigma_{pp}=6.6\times
10^{-26}\rm\: cm^{2}$ for CR protons with $E\sim 10^{15.5}$~eV
\cite{2006PhRvD..74c4018K}. Assuming that the protons are still confined
around the galactic center, the gamma-ray luminosity of molecular gas
around the center is given by
\begin{eqnarray}
 L_\gamma &\sim & f_{\rm MCe}L_{\rm p,tot}t_{\rm inj}/t_{pp}\nonumber \\
&=&6.8\times 10^{39}
\left(\frac{f_{\rm MCe}}{0.01}\right)
\left(\frac{L_{\rm p,tot}t_{\rm inj}}{3.5\times 10^{54}\rm\: erg}\right)
\left(\frac{t_{pp}}{1.6\times 10^5\rm\: yr}\right)^{-1}
\rm\: erg\: s^{-1}\:,
\end{eqnarray}
where $f_{\rm MCe}=f_{\rm MC} \min\{t_{\rm diff}/t_{pp}, 1\}$ is the
effective filling factor and $f_{\rm MC}$ is the filling factor of the
molecular gas. If the diffusion time-scale $t_{\rm diff}$ is smaller
than $t_{pp}$, a significant fraction of CR protons escape from the
molecular gas before they interact with protons in the gas. We
normalized $L_{\rm p,tot}t_{\rm inj}$ by our fiducial value. If the
distance to this galaxy is $d=30$~Mpc, the gamma-ray flux is
$f_\gamma=L_\gamma/(4\pi d^2)\sim 6.3\times 10^{-14}\rm\: erg\:
cm^{-2}\: s^{-1}$, which could be observed with LHASSO
\cite{2016CRPhy..17..663K}. If the flux is larger (e.g. larger $f_{\rm
MC}$ or smaller $d$), the galaxy could be detected with the Cherenkov
Telescope Array (CTA) \cite{2011ExA....32..193A}. Neutrinos are
additionally generated through $pp$-interaction. However, since the
expected flux is comparable to the gamma-ray flux, it would be difficult
to detect them in the near future. 

Gamma-rays are also created inside the RIAF, but high-energy gamma-rays
are absorbed by thermal photons there. Only photons of $E\lesssim 1$~GeV
can escape from the RIAF \cite{2015ApJ...806..159K}. The gamma-ray
luminosity is $\sim 0.1 L_{\rm p,tot}$ \cite{2015ApJ...806..159K}, which
means the flux of $\sim 2\times 10^{-13}\rm\: erg\: s^{-1}\: cm^{-2}$
for $d=30$~Mpc. This is marginally detectable by
Fermi\footnote{https://www.slac.stanford.edu/exp/glast/groups/canda/lat\_Performance.htm},
although gamma-rays from LLAGNs are hardly detected so far
\cite{2015A&A...584A..20W}. Neutrinos are also produced inside the RIAF,
and the neutrino flux is almost the same as the gamma-ray flux, which is
too low to be detected in the near future.

\section{Recent CR injection from Sgr~A*, and the gamma-rays and 
neutrinos from the CMZ}
\label{sec:CMZ}

In Paper~I, we indicated that the TeV gamma-rays observed at the CMZ are
generated by CRs recently ($\sim 10^5$~yr) injected by Sgr~A*. Since the
gamma-ray data have been updated with HESS \cite{2016Natur.531..476H},
we revisit our model here to show the consistency of our model with the
latest data.

We assume that the activities of Sgr~A* in the normal state, in which
Sgr~A* stays most of the time, are much weaker than those in the
outburst state, although they are much stronger than the activities in
the current faint state over the past $\sim 50$~yrs
\cite{1996PASJ...48..249K,2000ApJ...534..283M,2006PASJ...58..965T,2013PASJ...65...33R}. The
CR acceleration in the normal state may be much different from that in
the outburst state. The CRs responsible for the observed gamma-rays from
the CMZ may be accelerated during the normal state. Motivated by the
HESS observations that have shown that the CR spectrum in the CMZ is
described by a power-law \cite{2016Natur.531..476H}, we assume that the
spectrum of the CRs accelerated at Sgr~A* in the normal state is given
by eq.~(\ref{eq:spp}). The power-law spectrum may be realized when the
disk of the RIAF has a power-law structure and $E_{\rm p,eq}$ in
eq.~(\ref{eq:geq}) changes with the radius. Alternatively, CR
acceleration may occur at a vacuum gap in the black hole magnetosphere
\cite{2007ApJ...659.1063K,2016A&A...593A...8P}. Here, we chose
$\mu=4.07$, $p_{\rm S}c=9\times 10^{11}$~eV, and $p_{\rm L}c=1\times
10^{16}$~eV to be consistent with the gamma-ray observations. The CRs
are injected at the Galactic center ($r=0$). The normalization of
eq.~(\ref{eq:spp}) is given so that the energy injection rate is
$\lambda\eta_{\rm cr}\dot{M}c^2$, where $\lambda$ is the fraction
of CRs that enter the CMZ.  For the convenience to calculate gamma-ray
emission, we treat the diffusion of those CRs only in the direction of
the disk-like CMZ and we solve a spherically symmetric diffusion
equation,
\begin{equation}
\label{eq:diffcmz}
 \frac{\partial f_{\rm R}}{\partial t}
= \frac{1}{r^2}\frac{\partial}{\partial r}
\left(r^2 D_{\rm d}(p)\frac{\partial
f_{\rm R}}{\partial r}\right) + Q
\:,
\end{equation}
where $f_{\rm R}=f_{\rm R}(t,r,p)$ is the distribution function for the
recently injected CRs. The source term in Eq.~(\ref{eq:diffcmz})
is written as $\int 4\pi c p^3 Q dp = \lambda\eta_{\rm cr}
\dot{M}c^2$. Since the size of the CMZ ($\sim 100$~pc) is much larger
than that of the RIAF, we treat $Q$ as a point source. We consider
this component $f_{\rm R}$ only for $r\lesssim H_{\rm d}$ because those
CRs contribute little to the CRs in the halo at present (see
figure~\ref{fig:rad}).  We assume that the injection of those CRs is
steady and the left-hand side of eq.(\ref{eq:diffcmz}) is zero. The
diffusion coefficient in the CMZ is the same as that for the Galactic
disk $D_{\rm d}$ (eq.~\ref{eq:Dd}). In the following calculations, we
take $\eta_{\rm cr}=2\times 10^{-3}$, which is the same as that for the
outburst component (section~\ref{sec:CRacc}), and $\dot{m}=0.001$, which
is the same as that adopted in Paper~I. We also assume that
$\lambda=0.015$ to be consistent with observations. Since the diffusion
time of CRs in the CMZ is $\sim 10^4$~yr at $E\sim 10$~TeV, the results
do not much change even if $\dot{m}=0$ after the outburst $\sim 10^7$
years ago and before $\sim 10^4$~years ago. On the other hand, we assume
that the activities of Sgr~A* during this period is not too large to
affect the CRs observed on the Earth at present. The most recent
inactivity of Sgr~A* in the past $\sim 50$~years does not affect the
results and can be ignored.

Neutrinos and gamma-rays are produced via the interactions between the
CRs and target nucleons in the CMZ. The size of the CMZ is given by
$R_{\rm c}$ and the molecular gas is uniformly distributed for $r<R_{\rm
c}$. From Paper~I, we assume that $R_{\rm c}=130$~pc, and the gas
density of the CMZ is $\rho_{\rm c}=1.4\times 10^{-22}\rm\:
g\:cm^{-3}$. We calculate the production rate of gamma-ray photons and
neutrinos by pion decay using the formula provided by
Ref.~\cite{2006PhRvD..74c4018K}.

Using the results shown in figure~3 of Ref.~\cite{2006ApJ...640L.155M},
we estimate the attenuation of very high energy gamma-rays by pair
production on the Galactic interstellar radiation field. However, the
attenuation does not much affect the following results. The energy
density of interstellar radiation field ($\sim 10\rm\; eV\: cm^{-3}$) is
much smaller than probable magnetic fields in the CMZ, the gamma-ray
emission via inverse Compton scattering by secondary electrons can be
ignored \cite{2015PhRvD..92b3001F}.

\begin{figure}
\centering \includegraphics[width=90mm]{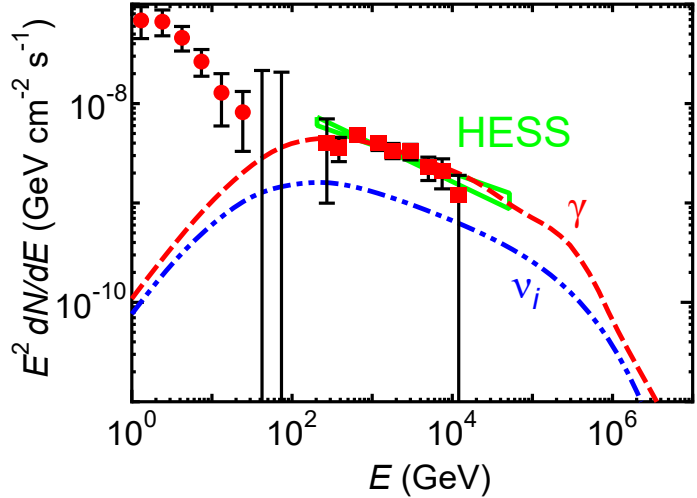}
\caption{Predicated
$\gamma$-ray flux (dashed line) and neutrino flux per flavor (two-dot
dashed line) from the CMZ. Filled circles and squares are the Fermi and
HESS observations, respectively
\cite{2013ApJ...762...33Y,2006Natur.439..695A}.  Recent HESS
observations (Ref.~\cite{2016Natur.531..476H}) are shown by the ribbon
assuming that the luminosity of the observed region is a factor of two
smaller than that of Ref~\cite{2006Natur.439..695A}.  \label{fig:cmz}}
\end{figure}

Figure~\ref{fig:rad} shows the radial distribution of this CR component,
which indicates that their influence is confined to the vicinity of the
Galactic center.  Figure~\ref{fig:cmz} shows the gamma-ray and neutrino
spectra of the CMZ. For $E\gtrsim 0.4$~TeV, the $\gamma$-ray spectrum is
consistent with the HESS observations. The spectrum extends even to
$E\gtrsim 10$~TeV, which is consistent with the recent report for the
inner region of the CMZ \cite{2016Natur.531..476H}. In our model,
associated 10-100~TeV neutrinos (figure~\ref{fig:cmz}) may be detectable
with KM3Net~\cite{2016arXiv160107459A}. The expected sensitivity is
$\sim{\rm a~few}\times{10}^{-9}~{\rm GeV}~{\rm cm}^{-2}~{\rm s}^{-1}$,
so the detection would be feasible with many years of observations. The
$\gamma$-rays from the CMZ at $E\sim 100$~TeV (figure~\ref{fig:cmz}) may
be detected with CTA \cite{2011ExA....32..193A}. Note that the
point-like gamma-ray source HESS J1745--290 may be the direct emission
from Sgr~A* and may reflect its current activity.

\section{Power-law CR spectrum at the outburst of Sgr~A*}
\label{sec:power}

In the fiducial model in section \ref{sec:result}, we take a typical
spectrum of stochastic acceleration for the CR spectrum at the outburst
of Sgr~A* (eq.~\ref{eq:sp}).  Here, we consider a power-law spectrum for
the outburst, because the CR spectrum for the recent injection is a
power-law (section~\ref{sec:CMZ}).

We calculate the CR spectrum on the Earth when the CR spectrum at the
outburst is given by eq.~(\ref{eq:fint}) in which $\dot{N}$ is replaced
by the power-law form (eq.~\ref{eq:spp}). The total energy injection
rate of the CRs is given by $L_{\rm p, tot}t_{\rm inj}$
(section~\ref{sec:diff}) and the value is the same as that for the
fiducial model ($\dot{m}=0.1$, $\eta_{\rm cr}=2\times 10^{-3}$, and
$t_{\rm inj}=1$~Myr). The cutoff momentums ($p_{\rm S}c=9\times
10^{11}$~eV, and $p_{\rm L}c=1\times 10^{16}$~eV) are the same as those
for the recently injected component that can explain the gamma-ray data
observed with HESS (section~\ref{sec:CMZ}). The lower-cutoff may be
determined by the advection of CRs by some outflows. Moreover, we set
$x=p/p_{\rm S}$ in eq.~(\ref{eq:fint}). As is the case of the fiducial
model, we assume that $r_{\rm s}=0.1$~kpc, and the initial distribution
function for $r>r_{\rm s}$ is zero.

\begin{figure}[tbp]
\centering \includegraphics[width=90mm]{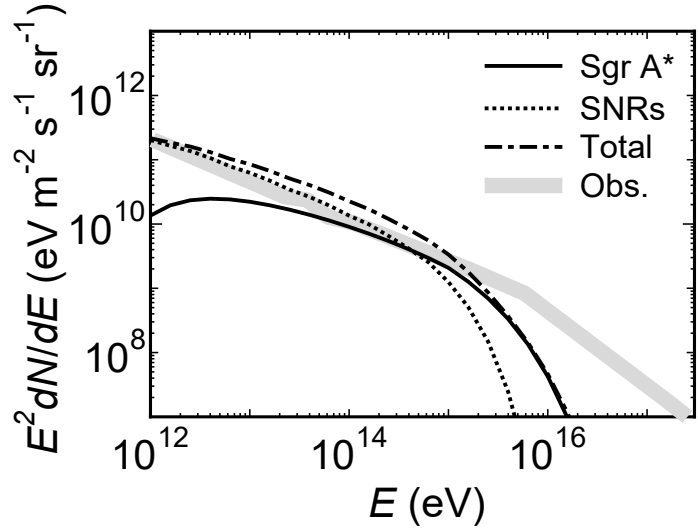}
\caption{\label{fig:power} Same as figure~\ref{fig:sp} but for the
power-law injection (eq.~\ref{eq:spp}) similar to the recent injection
(figure~\ref{fig:cmz}).}
\end{figure}

In figure~\ref{fig:power}, we show the CR spectra on the Earth at
$t=10$~Myr. It shows that the CR flux is almost comparable to that from
the SNRs and suggests that Sgr~A* alone could provide enough CRs that
are detected on the Earth. However, the slope of the spectrum does not
match the observation and the spectrum is curved. This is because the CR
injection from Sgr~A* is instantaneous; higher-energy CRs have already
arrived on the Earth while most of lower-energy CRs have not, which is
the same reason for the curve of the anisotropy plot
(figure~\ref{fig:aniso}). Moreover, the highest-energy CRs ($E\gtrsim
10^{15}$~eV) have started to escape from the Galactic halo, which
decreases the number density of those CRs on the Earth. These facts
suggest that it would be difficult to explain the CR spectrum on the
Earth only by those accelerated by Sgr~A* with the same spectral form
observed at the CMZ. To be consistent with both the latest HESS data for
the CMZ and the CR flux on the Earth, the acceleration mechanism at the
outburst of Sgr~A* may need to be somewhat different from that for the
recent injection.

\section{Conclusions}
\label{sec:conc}

We have shown that a past intense activity of Sgr~A*, which created the
Fermi bubbles $\sim 10$~Myr ago, can significantly contribute as a
``Pevatron'' to the Galactic CRs around the knee ($E\sim 10^{15.5}$~eV)
observed on the Earth. The diffusion coefficient in the halo is
estimated on the condition that the CRs are prevailing in the halo at
present. We solved a diffusion equation for the CRs and reproduced the
observed CR flux around the knee. The observed small anisotropy of the
arrival directions of the CRs is compatible with the prediction if the
diffusion coefficient in the Galactic disk is smaller than that in the
halo, which is reasonable. Our model predicts that the boron-to-carbon
ratio is independent of CR energy at energies close to the knee if the
CRs from Sgr~A* are dominant around that energy. Gamma-ray emissions
could be observed in nearby galaxies, if similar activities are
happening there. It is unlikely that the spectrum of the CRs accelerated
at the outburst of Sgr~A* is described by a power-law with a large
energy range in contrast with the CR spectrum suggested from the
gamma-ray observations of the CMZ. This may mean that the CR
acceleration mechanism at the outburst could be different from that in
the normal state of Sgr~A*.

Although our single-source scenario may be rather extreme and some
tuning is necessary to fit the observed CR flux smoothly, surprisingly
such models are not ruled out by the present data. Thus, our results
imply that Sgr A* should be examined as one of the potential sources of
CRs around the knee, whose origin has been discussed for many
years~(e.g.,~\cite{2003APh....19..193H,2004APh....21..241H}). We still
cannot deny the possibility that SNRs are still the main sources for the
CRs around the knee, and proposed ideas including faster acceleration at
oblique shocks~\cite{2002PhRvD..66h3004K,2011MNRAS.418.1208B},
contributions from Type IIn and Type IIb
supernovae~\cite{2003A&A...409..799S,2014MNRAS.440.2528M,2010ApJ...718...31P},
and early acceleration at dense stellar
winds~\cite{1993A&A...274..902S,2014MNRAS.440.2528M,2013MNRAS.435.1174S}.
A different source population such as super-bubbles has also been
considered~(e.g.,~\cite{2015MNRAS.453..113B,2016PhRvD..93h3001O}). However,
our study motivates investigations into the roles of the Galactic center
in CR production, possibilities of a single source origin, and so
on~\cite{2001JPhG...27.1005E}.  We note that models with a recent
local source may give results similar to ours. If the source exploded
near the Earth and the PeV CRs have already diffused out on a scale much
larger than the distance to the source, the anisotropy of the arrival
directions can be small. However, a problem of these models is that the
same sources are likely to be widely distributed in the Galaxy
especially if the PeV CR source is an SNR. This encounters the
difficulty mentioned in section~\ref{sec:intro} that there is little
observational evidence implicating SNRs as Pevatrons.

\appendix
\section{Analytical solutions for the diffusion equation}

If the Galactic disk can be ignored, the diffusion of CRs in the
spherical halo can be analytically solved. The diffusion equation is
\begin{equation}
\label{eq:diffA}
 \frac{\partial f}{\partial t}
= \frac{1}{r^2}\frac{\partial}{\partial r}
\left(r^2 D_{\rm h}(p)\frac{\partial
					       f}{\partial r}\right)
\:,
\end{equation}
where $f=f(t,r,p)$ is the distribution function. We assume that $f=0$ at
$r=R_{\rm h}$ and $\partial f/\partial r=0$ at $r=0$. In
eq.~(\ref{eq:diffA}), $D_{\rm h}$ can be treated as a constant for a
given $p$. Using the separation of variable method, eq.~(\ref{eq:diff})
is represented as
\begin{equation}
 \frac{1}{r^2R}\frac{d}{dr}\left(r^2\frac{dR}{dr}\right)
= \frac{1}{D_{\rm h} T}\frac{\partial T}{\partial t}
\end{equation}
for $f(t,r,p)=R(r)T(t,p)$. This equation makes sense when both sides are
equal to a negative constant $-\xi^2$. Thus, we obtain two equations:
\begin{equation}
\label{eq:T}
 \frac{\partial T}{\partial t} + \xi^2 D_{\rm h} T = 0\:,
\end{equation}
\begin{equation}
\label{eq:R}
 \frac{1}{r^2}\frac{d}{dr}\left(r^2\frac{dR}{dr}\right)
+ \xi^2 R = 0.
\end{equation}
The solution for eq.~(\ref{eq:T}) is
\begin{equation}
 T(t,p) = A \exp(-D_{\rm h}(p)\xi^2 t)\:,
\end{equation}
where $A$ is the constant. 
The solution of eq.~(\ref{eq:R}) is
\begin{equation}
 R(r) = B'\frac{\sin(\xi r)}{r} + C'\frac{\cos(\xi r)}{r}
= B'\frac{\sin(\xi r)}{r}
\:,
\end{equation}
where $B'$ and $C'$ are the constants, and $C'$ must be zero because $f$
is finite at $r=0$. Because of the boundary condition at $r=R_{\rm h}$,
the eigenvalue satisfies
\begin{equation}
 \xi_n = n\pi/R_{\rm h}
\end{equation}
and the eigenfunctions are
\begin{equation}
 R_n(r) = B'_n\frac{\sin(\xi_n r)}{r}
\end{equation}
for $n=1,2,3, ...$.

Thus, the solution for the number $n$ is
\begin{equation}
 f_n(t,r,p)=R_n(r)T(t,p) = \frac{A_n}{r}\sin(\xi_n r)
\exp(-D_{\rm h}(p)\xi_n^2 t)\:,
\end{equation}
where $A_n = A B'_n$. The general solution is given by their
superposition:
\begin{equation}
 f(t,r,p)=\sum_{n=1}^\infty f_n(t,r,p)
= \sum_{n=1}^\infty \frac{A_n}{r}\sin(\xi_n r)
\exp(-D_{\rm h}(p)\xi_n^2 t)\:.
\end{equation}
The coefficients $A_n$ can be determined by the initial condition at
$t=0$:
\begin{equation}
 f(0,r,p) = \sum_{n=1}^\infty \frac{A_n}{r}\sin(\xi_n r)\:.
\end{equation}
Since this is a Fourier series, the coefficients are given by
\begin{equation}
 A_n(p) = \frac{2}{R_{\rm h}}\int_0^{R_{\rm h}}f(0,r',p)
\sin(\xi_n r')r' dr'\:.
\end{equation}

\acknowledgments

This work was supported by MEXT KAKENHI No. 15K05080 (YF). The work of
K.M. is supported by NSF Grant No. PHY-1620777 (KM). This work is partly
supported by NASA NNX13AH50G and by IGC post-doctoral fellowship program
(S.S.K).

\bibliographystyle{JHEP}
%\bibliography{sgras}
\providecommand{\href}[2]{#2}\begingroup\raggedright\endgroup

\end{document}